\documentclass{aa}
\usepackage{natbib}
\usepackage{graphicx}

\newcommand{\gl}{$\lambda$}

\begin{document}

\title{The EUVE point of view of AD Leo}
\author{J. Sanz-Forcada, \and G. Micela}
\institute{INAF - Osservatorio Astronomico di Palermo
G. S. Vaiana, Piazza del Parlamento, 1; Palermo, I-90134, Italy}
\offprints{J. Sanz-Forcada, \email{jsanz@astropa.unipa.it}}
\date{Received / Accepted}

\abstract{
All the {\em Extreme Ultraviolet Explorer} (EUVE) observations of
AD~Leo, totalling  1.1~Ms of exposure time, have been employed 
to analyze the corona of this single M dwarf.
The light curves show a
well defined quiescent stage, and a distribution of amplitude of
variability following a power law with a $\sim -2.4$ index. 
The flaring behavior exhibits much similarity with other M
active stars like FK~Aqr or YY~Gem, and flares behave differently from 
late type active giants and subgiants.  
The Emission Measure Distribution
(EMD) of the summed spectrum, as well as that of quiescent and flaring
stages, were obtained using a line-based method. The average EMD is
dominated by material at log~T(K)$\sim$6.9, with a second peak around
log~T(K)$\sim$6.3, and a large increase in
the amount of material with log~T(K)$\ga$7.1 during flares, material
almost absent during quiescence. 
The results are interpreted as the combination of three families of
loops with maximum temperatures at log~T(K)$\sim$6.3, $\sim$6.9 and
somewhere beyond log~T(K)$\ga$7.1. 
A value of the abundance of [Ne/Fe]=1.05$\pm$0.08 was measured at
log~T(K)$\sim$5.9. No significative increment of Neon abundance was
detected between quiescence and flaring states.

}

\maketitle

\section{Introduction}

AD Leo (GL 388) is a dM3 star located at a distance of 4.9~pc \citep{hen94}.
It is a well known frequent source of 
flares. Its high activity is probably due to its high rotation
rate \citep[P$_{\rm phot}\sim$2.7~d,][]{spi86}, 
and a large number of studies on this star has been carried out 
because of its nature as a very active single star. 
M stars are supposed to have a large convective layer, resulting in a
high level of photospheric spot coverage and frequent flaring
activity. With a mass of $M = 0.40 M_{\odot}$, \object{AD Leo}
would still be close to the limit where the radiative core 
of the star is still present \citep{fav00}. Studies in X-rays reflect
that M stars are the stellar class 
with highest L$_X$/L$_{\rm bol}$ values, binary systems or very young
stars apart. 

AD~Leo is one of the paradigms of flaring stars, and has been subject
of frequent studies in the EUV and X-rays.
X-rays low resolution spectra have been used by several authors
\citep[see][ and references therein]{fav00} to obtain fits to 2 or 3
temperatures that could explain such 
spectra.  With the advent of the {\em Extreme Ultraviolet Explorer}
(EUVE) it was possible to obtain the first  
high resolution spectra, and an analysis based on 85~ks of observation
permitted \citet{cul97} to get an Emission
Measure Distribution (EMD) during different activity levels. 
However, the low statistics of
the observation did not allow an accurate analysis of the EMD to be performed,
and in the best
case only 3 lines had signal-to-noise ratio (S/N) higher than 4, making only
indicative the analysis 
on the variations between quiescent and flaring stages. Some
studies relative to the EUVE light curves have been carried out by
\citet{haw95}, \citet{gud01}, and \citet{haw01}, but no deep analysis in has
been conducted from high quality EUVE spectra to date. 

More recently, the analysis of Chandra/Low-Energy Transmission Grating
(LETG) spectrum \citep{mag01} 
has been used to get values of the abundances, and an EMD peaking at
log~T(K)$\sim$6.8. 
The EUVE satellite (\gl\gl$\sim$70--750~\AA) has the advantage
in respect of Chandra (\gl\gl$\sim$1--175) and 
the {\em X-ray Multi Mirror telescope} (XMM-Newton)
(\gl\gl$\sim$5--35) of a good 
coverage of lines of just one element. Iron lines observed with EUVE permit
calculating the structure of the EMD in the range
log~T(K)$\sim$5.8--7.4 without any ambiguity produced by the use of
different values of abundances of the elements observed.  In contrast,
Chandra or XMM-Newton can observe a larger number of lines, also with higher
spectral resolution, from different
elements, covering a larger temperature region [up to
log~T(K)$\sim$8]. But in order to get a good temperature coverage, the
spectral lines corresponding to different elements must be used. Hence 
if no proper calculation of the abundances is performed in the analysis of
such spectra, this can result in a wrong EMD.  While the combination
of the observations with EUVE and Chandra or XMM-Newton is desirable, the
frequent variability of a star like AD~Leo makes such analysis difficult. 
Finally, the observations made with EUVE tend to span long periods of
time (several days) in order to achieve good statistics in the
spectra, permitting to get a good analysis on coronal variability, and
even a separate analysis of flaring and quiescent stages, as it has
been done, for example, for several RS~CVn stars \citep{sanz01,paper1}.  

\begin{table}
\caption{\scshape EUVE exposure times (s) of the AD~Leo observations (SW
  and MW spectra)}\label{times} 
\begin{center} 
\begin{tabular}{lrr}
\hline \hline
{Start date}  & {SW}  & {MW} \\
\hline
{1 Mar 1993}  & {84\,622}  & 85\,271 \\
{3 Mar 1996}  & {73\,322}  & 73\,320 \\
{2 Apr 1999}  & {46\,441}  & 44\,384 \\
{5 Apr 1999}  & {141\,885} & 141\,888 \\
{9 Apr 1999}  & {133\,001} & 131\,198 \\
{17 Apr 1999} & {177\,093} & 176\,239 \\
{25 Apr 1999} & {158\,084} & 145\,181 \\
{6 May 1999}  & {190\,759} & 147\,785 \\
{9 Mar 2000}  & {99\,952}  & 52\,686 \\
\hline
\end{tabular}
\end{center}
\end{table}

A total of 1~Ms of observations converts AD Leo in the active star
most observed with EUVE, yielding 
a combined spectrum
with high statistics, and it permits to study the 
variability properties 
observed during an elapsed time of 46~d.  In this paper we are presenting
the most accurate analysis to date on the EMD of an M star without
the ambiguity derived from the use of different elements. The long
duration of the observations of AD~Leo allows us to perform an analysis
of the flaring behavior from high quality spectra, and from the light
curves as compared to the
Sun. Given the spectral range response of 
the EUVE/DS ($\lambda\lambda$70--175~\AA), such light
curve will be dominated by emitting material with log~T(K)$\sim$6.7--7.2.

A description of the technical
information of the observations is given in
Sect.~\ref{sec:observations}. The analysis of data (light curves,
spectra, and Emission Measure Distribution) is
described in Sect.~\ref{sec:analysis}. Results are discussed in
Sect.~\ref{sec:discussion}
in the context of coronal activity in this and other late type stars,
followed by a summary of the conclusions of the work
(Sect.~\ref{sec:conclusions}). 

\section{Observations}\label{sec:observations}
A total of 1~Ms of EUVE observations of AD~Leo taken between 1993 and
2000 (Table~\ref{times}) were made available through the EUVE Data
Archive. The EUVE spectrographs cover the
spectral range 70--180~\AA, 
170--370~\AA\  and 300--750~\AA\  for the short-wavelength (SW),
medium-wavelength (MW) and long-wavelength (LW)  spectrometers
respectively, with corresponding spectral dispersion of
$\Delta\lambda\sim$ 0.067, 0.135, and 0.270~\AA/pixel, and an
effective spectral resolution of
$\lambda/\Delta\lambda$$\sim$200--400. The Deep Survey (DS) Imager has
a bandpass of 80--180~\AA\ and is used for EUV photometry.

\begin{table}
\caption{EUVE line fluxes in AD Leo$^a$}\label{euveflux}
\tabcolsep 3.pt
\begin{tabular}{lrrcrcrc}
\hline \hline
 {}& {$\lambda$$_{lab}$} & \multicolumn{2}{c}{Summed} &
 \multicolumn{2}{c}{Quiescent} & \multicolumn{2}{c}{Flaring}  \\
\cline{3-8} 
{Ion} & {(\AA)} & {S/N} & {Flux} & {S/N} & {Flux} & {S/N} & {Flux} \\
\hline
\multicolumn{8}{l}{\em Short Wavelength Spectrometer} \\
\ion{Ne}{viii}$^b$ &  88.08 &  12.9 & 1.47e-04 &  10.1 & 1.09e-04 &   5.8 & 2.13e-04 \\
\ion{Fe}{xix}$^c$ &  91.02 &   8.2 & 7.42e-05 & $\cdots$ & $\cdots$ &   3.9 & 1.21e-04 \\
\ion{Fe}{xviii} &  93.92 &  22.8 & 3.32e-04 &  19.7 & 2.90e-04 &  10.6 & 4.75e-04 \\
\ion{Ne}{viii}$^d$  &  98.26 &  20.6 & 2.86e-04 &  17.9 & 2.47e-04 &  10.7 & 5.24e-04 \\
\ion{Fe}{xix} & 101.55 &   6.4 & 5.30e-05 &   6.6 & 4.80e-05 &   4.3 & 1.06e-04 \\
\ion{Fe}{xxi} & 102.22 &  13.6 & 1.39e-04 &  11.4 & 1.13e-04 &   8.1 & 3.22e-04 \\
\ion{Ne}{viii}  & 103.08 &   8.5 & 7.20e-05 &   7.7 & 6.29e-05 & $\cdots$ & $\cdots$ \\
\ion{Fe}{xviii} & 103.94 &  11.0 & 1.05e-04 &  10.0 & 8.97e-05 &   5.0 & 1.27e-04 \\
\ion{Fe}{xix} & 106.33 &   5.6 & 3.62e-05 &   5.6 & 3.81e-05 &   3.3 & 7.15e-05 \\
\ion{Fe}{xix}$^e$ & 108.37 &  19.3 & 2.37e-04 &  15.5 & 1.83e-04 &  10.9 & 4.91e-04 \\
\ion{Fe}{xix} & 109.97 &   5.3 & 3.28e-05 &   4.0 & 2.39e-05 &   3.1 & 6.20e-05 \\
\ion{Fe}{xx} & 110.63 &   4.4 & 3.08e-05 &   3.4 & 2.05e-05 &   3.4 & 7.24e-05 \\
\ion{Fe}{xix} & 111.70 &   5.0 & 3.86e-05 &   4.0 & 2.65e-05 &   4.1 & 1.02e-04 \\
\ion{Fe}{xxii} & 114.41 &   7.7 & 6.47e-05 &   5.0 & 4.12e-05 &   4.6 & 1.22e-04 \\
\ion{Fe}{xxii}$^f$ & 117.17 &  19.4 & 2.76e-04 &  13.2 & 1.78e-04 &  11.1 & 6.05e-04 \\
\ion{Fe}{xx} & 118.66 &   8.3 & 7.42e-05 &   6.9 & 6.61e-05 &   5.4 & 1.92e-04 \\
\ion{Fe}{xix} & 120.00 &   7.6 & 7.57e-05 &   8.2 & 8.45e-05 &   4.0 & 1.20e-04 \\
\ion{Fe}{xxi} & 121.21 & $\cdots$ & $\cdots$ & $\cdots$ & $\cdots$ &   2.1 & 2.68e-05 \\
\ion{Fe}{xx} & 121.83 &  13.9 & 1.82e-04 &  10.0 & 1.20e-04 &  10.3 & 5.72e-04 \\
\ion{Fe}{xxi} & 128.73 &  17.0 & 2.90e-04 &  13.2 & 2.00e-04 &  12.3 & 8.94e-04 \\
\ion{Fe}{xxiii}$^g$ & 132.85 &  30.4 & 8.69e-04 &  23.2 & 5.95e-04 &  19.9 & 2.40e-03 \\
\ion{Fe}{xxii} & 135.78 &  14.0 & 2.25e-04 &  11.0 & 1.67e-04 &   8.4 & 5.18e-04 \\
\ion{Fe}{xxi}$^h$ & 142.16 &   7.7 & 1.00e-04 &   5.4 & 6.45e-05 & $\cdots$ & $\cdots$ \\
\ion{Fe}{ix}$^i$ & 171.07 &   9.5 & 3.26e-04 &   8.5 & 3.08e-04 &   5.0 & 3.19e-04 \\
\ion{Fe}{x}$^i$ & 174.53 &   8.2 & 3.31e-04 &   8.0 & 3.45e-04 & $\cdots$ & $\cdots$ \\
\ion{Fe}{xi}$^{i,j}$ & 180.40 & $^*  4.1$ & 1.52e-04 & $\cdots$ & $\cdots$ & $\cdots$ & $\cdots$ \\
\multicolumn{ 8}{l}{\em Medium Wavelength Spectrometer} \\
\ion{Fe}{ix}$^i$ & 171.07 & $^*  8.6$ & 2.50e-04 & $^*  8.9$ & 2.83e-04 & $^*  3.0$ & 2.40e-04 \\
\ion{Fe}{x}$^i$ & 174.53 & $\cdots$ & $\cdots$ & $\cdots$ & $\cdots$ &   3.9 & 2.83e-04 \\
\ion{Fe}{x}$^i$ & 177.24 & $\cdots$ & $\cdots$ & $\cdots$ & $\cdots$ &   4.1 & 3.42e-04 \\
\ion{Fe}{xi}$^{i,j}$ & 180.40 &   5.9 & 1.31e-04 &   5.9 & 1.34e-04 &   4.8 & 3.57e-04 \\
\ion{Fe}{xxiv}$^k$ & 192.04 & $\cdots$ & $\cdots$ & $\cdots$ & $\cdots$ &  10.3 & 1.65e-03 \\
\ion{Fe}{xii}$^l$ & 193.51 &  18.9 & 1.07e-03 &  16.0 & 9.00e-04 &   5.0 & 4.18e-04 \\
\ion{Fe}{xii} & 195.12 & $\cdots$ & $\cdots$ & $\cdots$ & $\cdots$ &   4.4 & 3.66e-04 \\
\ion{Fe}{xiii}$^m$ & 203.83 &   9.0 & 2.69e-04 &   8.9 & 2.76e-04 &   3.5 & 2.81e-04 \\
\ion{Fe}{xiv$^n$} & 211.33 &  10.1 & 3.16e-04 &   7.7 & 2.25e-04 &   3.3 & 2.58e-04 \\
\ion{He}{ii} & 256.32 &  15.8 & 8.38e-04 &  13.6 & 7.29e-04 &   9.6 & 1.94e-03 \\
\ion{Fe}{xv} & 284.15 &  15.6 & 6.87e-04 &  13.4 & 6.28e-04 &   7.3 & 1.05e-03 \\
\ion{Fe}{xvi} & 335.41 &  12.5 & 3.99e-04 &  12.4 & 4.42e-04 &   5.1 & 3.83e-04 \\
\ion{Fe}{xvi} & 360.80 &   6.5 & 1.14e-04 &   5.9 & 1.10e-04 & $\cdots$ & $\cdots$ \\
\hline
\end{tabular}

\begin{scriptsize}
$^a$``Flux'' represents  Flux at Earth,
  expressed in ph\,cm$^{-2}$\,s$^{-1}$ in the line. Iron lines marked with *
  were not considered in the EMD calculation. Columns with S/N
  represent the signal-to-noise ratio (see text).       

$^b$Blend with \ion{Ne}{viii} \gl 88.12. Both lines are included in
  the measurement.

$^c$Blend with \ion{Fe}{xxi} $\lambda$91.28. Both lines
  are included in measurement.

$^d$Blend with \ion{Fe}{xxi} \gl 97.88, \ion{Ne}{viii} \gl 98.11,
  \gl 98.27. All lines are included in measurement and modeled
  accordingly. 

$^e$Blend with \ion{Fe}{xxi} \gl 108.12. Both lines are included in
  the measurement.

$^f$Blend with \ion{Fe}{xxi} \gl 117.51. Both lines are included in
  the measurement.

$^g$Blend with \ion{Fe}{xx} $\lambda$132.85.  Both lines are 
included in measurement and modeled accordingly.

$^h$Blended with  \ion{Fe}{xxi} $\lambda$142.27. Both
  lines are included in measurement.

$^i$These lines are near the spectrometer limits and may
 be difficult to measure in either SW or MW. The lack of redundant
 measurements indicates that the lines were weak and/or noisy.

$^j$Blend with \ion{Fe}{x} \gl 180.41, \ion{Fe}{xxi} \gl 180.55 and
\ion{Fe}{xi} \gl 180.60. All lines  included in measurement and
modeled accordingly. 

$^k$May include blend of \ion{O}{v}
($\lambda$192.751, $\lambda$192.799, $\lambda$192.906),
\ion{Fe}{xii} ($\lambda$192.394) and possibly other weaker components.

$^l$Blend with \ion{Fe}{xxiv} \gl 192.04 (and its mentioned blends)
and \ion{Fe}{xii} \gl 195.12, in
quiescent and summed stages, are included in the measurement. 

$^m$Complex blend with \ion{Fe}{xiii} \gl 203.79, \gl 204.26,
\ion{Fe}{xvii} \gl 204.65, and possibly other weaker components. 

$^n$Blend with \ion{Fe}{xii} \gl 211.74. Both lines are included in
measurement. 

\end{scriptsize}
\end{table}

\begin{figure*}
  \includegraphics[width=\textwidth]{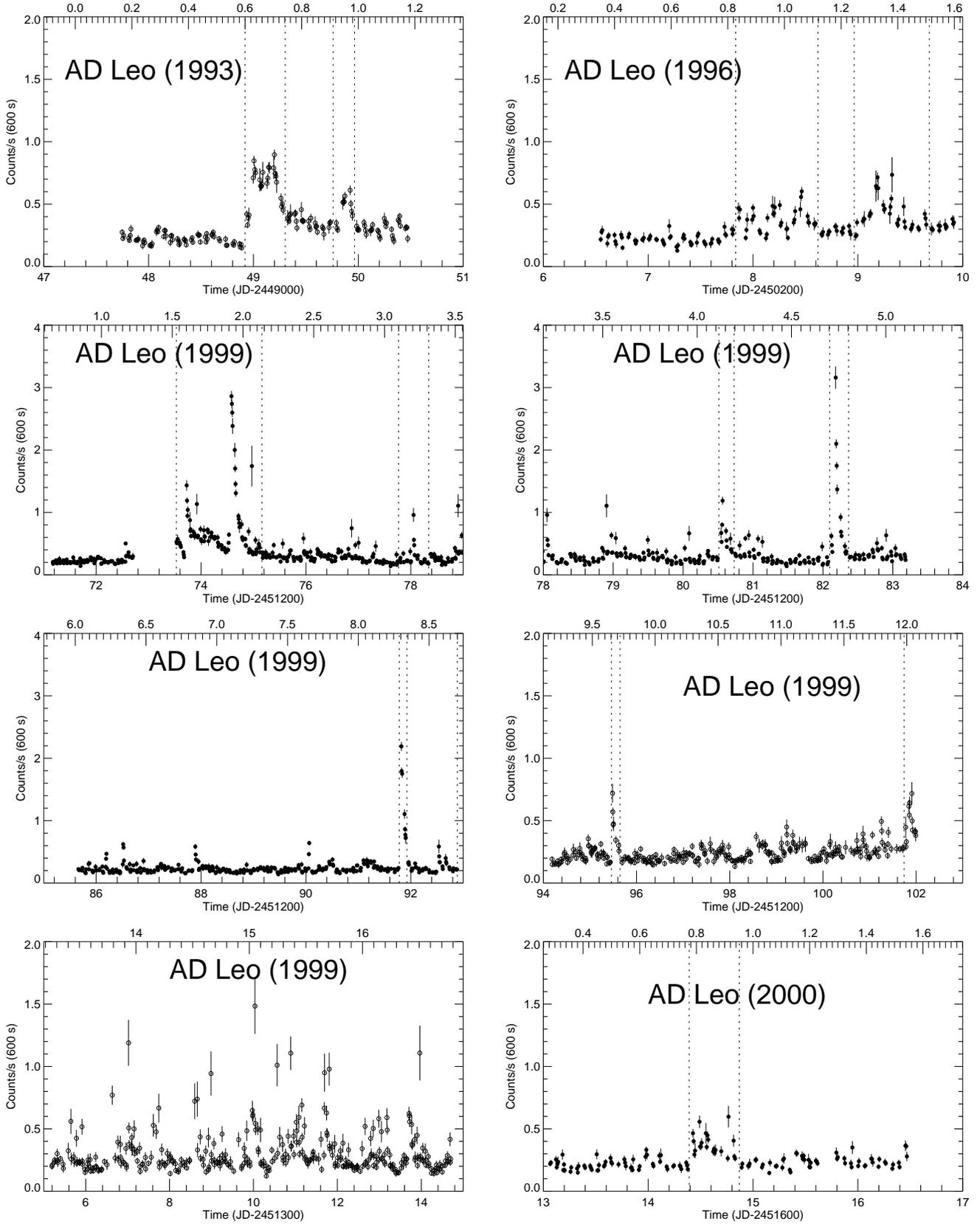}  
  \caption{DS light curves as a function of Julian Date (lower axis)
  and rotational phase (upper axis), based on a P=2.7~d \citep{spi86}.
  Rotational phase 0 is arbitrary. Open circles are data affected
  by the dead spot, while solid circles represent unaffected
  data. 1-$\sigma$ error bars are shown. Only points with S/N
  higher than 5 are plotted. Vertical dashed lines mark the
  separation between quiescent and flaring stages. The bin size is
  600~s.}
  \label{lcurve}
\end{figure*}

\begin{figure*}
  \includegraphics[width=8.4cm]{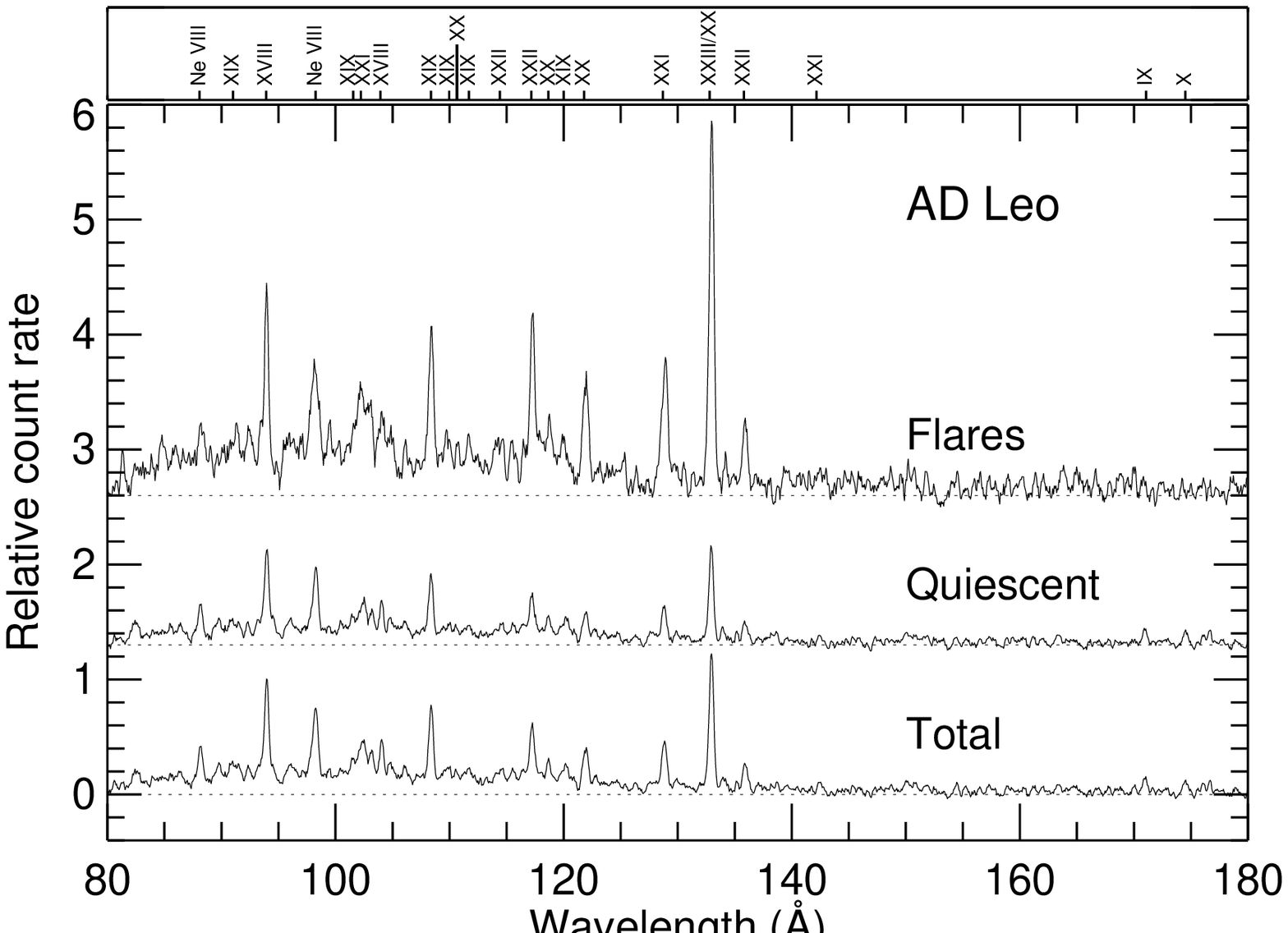}
  \includegraphics[width=8.4cm]{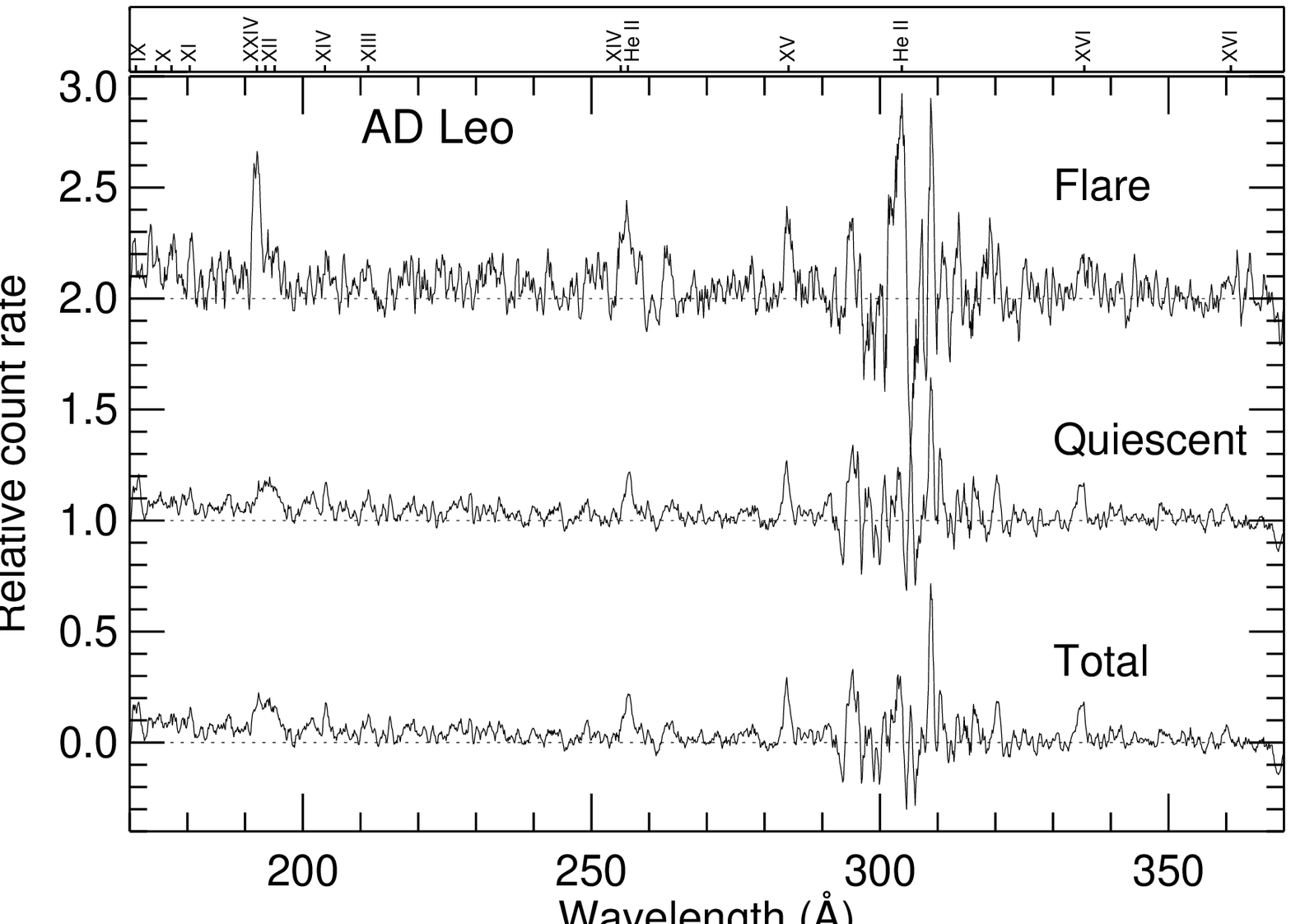}
  \caption{EUVE SW and MW spectra for AD Leo during the flare and
  quiescent intervals, and the summed (marked Total) spectra. Ion
  stages of iron are marked in the top panel. Spectra are smoothed by
  5 pixels. Spectra for each target have been normalized to the same
  exposure time and then offset for display.  
  Dotted lines indicate the zero flux level of each
  spectrum.} 
  \label{adleospec}
\end{figure*}

EUVE light curves for the target (Fig.~\ref{lcurve}), were made from
the DS image, by taking a circle centered on the source, and
subtracting the equivalent sky background within an annulus measured around
the center. Standard procedures were used in the IRAF package EUV
v1.9, with a time binning of 600~s. Points affected by the ``dead
spot'' are marked as open circles in the light curves, while filled circles
mark the unaffected points. The ``dead spot'' is a low gain 
area of the DS detector that affects some of the EUVE observations 
taken in 1993 and 1994, resulting in variable levels of contamination 
of the signal \citep[see][]{mill95}. However, part of the observations
of AD~Leo taken in 1999 seem to be partially affected by the
``dead spot''. Although the flux seems to be at the same level as the
rest of the 1999 campaign, the affected points (both in
1993 and in 1999) were discarded for the
statistical analysis on variability of the star. 

The variations observed in the light curves show different activity
levels, including some flaring episodes. Quiescent and flaring states
were identified in the light curves; vertical lines separate the
different states in the light curves. 
Spectra binned over these selected intervals were extracted from the
SW and MW spectrographs (Fig.~\ref{adleospec}), but since the LW spectra did
not show enough statistics for its 
extraction in all intervals, they were discarded. 
Line fluxes of strong lines identified in the summed, flaring and
quiescent EUVE spectra, 
are given in Table~\ref{euveflux}. S/N of the
lines is defined as $S/[S+B(1+1/n)]^{1/2}$, where S is net signal, B is the
estimated average background, and n is the oversampling ratio
(i.e., the number of background pixels to the number of source
pixels in the image), having a typical value of n$\sim$10--15 in
our extraction.  S and B are calculated for the total integrated
line signal (minus continuum for SW lines) and background.

\section{Data Analysis}\label{sec:analysis}   

\subsection{Light Curves}

The EUVE light curves of AD~Leo show the presence of frequent flaring
events, usually of ``short'' duration (less than 1 day) but quite
intense (Fig.~\ref{lcurve}). Although in the case of the 1993 campaign
a flaring event starting at JD 2,449,049 was interpreted as only
one flare by \citet{haw95}, the complex structure observed
might actually correspond to the overlap of two or even three flares
happening close in time. The observed behaviour in this flaring event
does not seem to have repeated in the rest of the AD~Leo observations. 
The presence of short and intense flares in the EUV and
X-rays light curves is usually found in other dM stars, such as 
\object{FK Aqr} or \object{YY Gem} \citep{paper2}, and it contrasts
with the long duration flares found for 
some active binary systems with an evolved component \citep{paper1}, 
like UX~Ari (K0IV/G5V), $\sigma$~Gem (K1III/?) or V711~Tau (K1IV/G5IV).
Moreover, the short duration flares seem to be
common among dwarf stars of earlier types too, like LQ~Hya (K2V) or
$\sigma^2$~CrB (G0V/F6V), either in EUVE or in X-rays \citep[see, for
instance,][]{cov01,ost00,paper2}. 

\begin{figure*}
  \includegraphics[width=8.4cm]{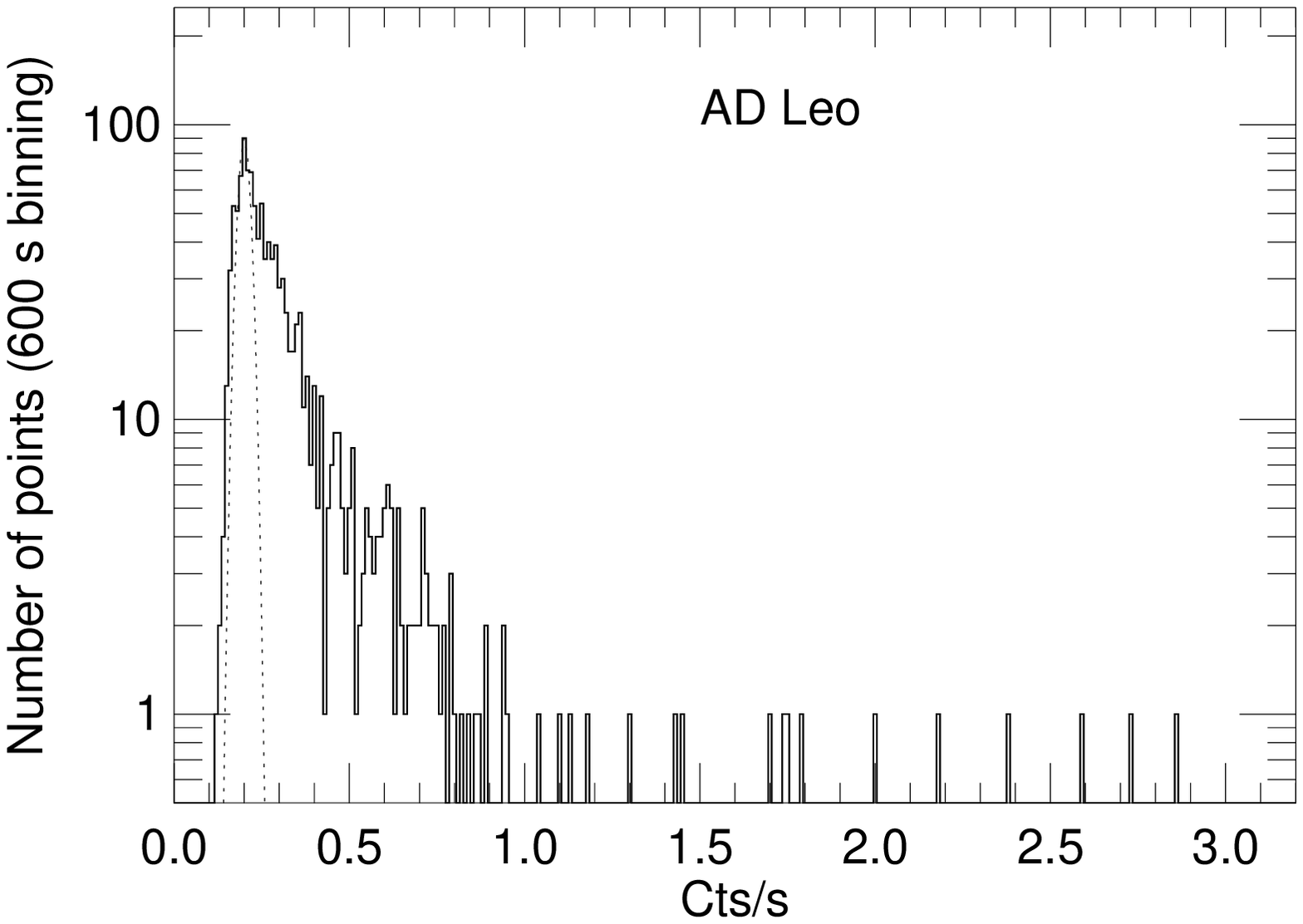}
  \includegraphics[width=8.4cm]{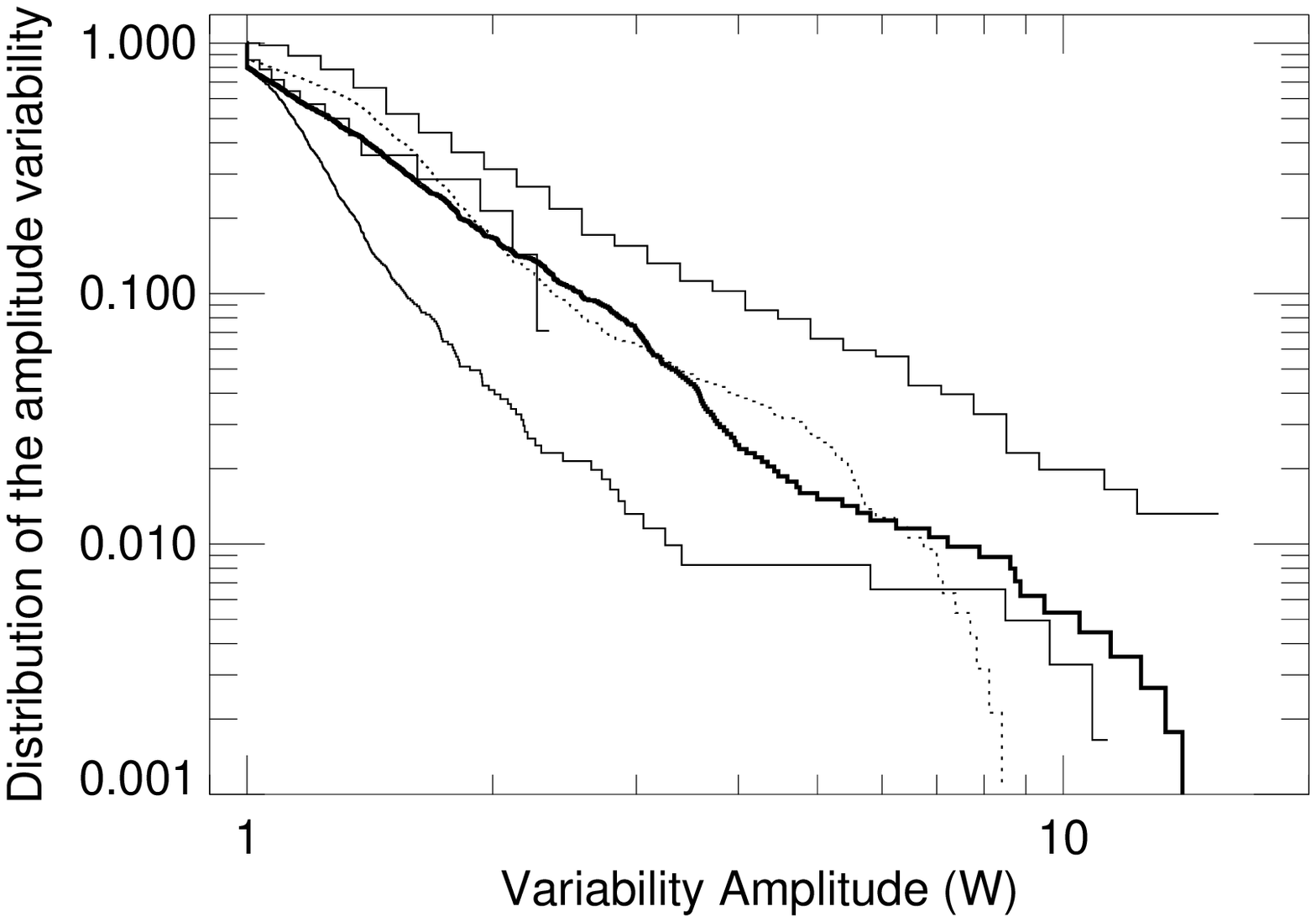}
  \caption{{\em Left:} EUV (80--180~\AA) luminosity distribution of
    variability for a 600~s binning. The thin dotted line represents
    the fluctuations distribution of the quiescent EUV level (L$_{\rm qsc}$,
    see text). {\em Right:} Normalized cumulative distribution of the
    amplitude (W=L$_{\rm EUV}$/L$_{\rm min}$) variability in AD~Leo (thick
    line). Lower thin line and dotted line represent the distribution
    of YY~Gem and FK~Aqr from EUVE observations \citep{paper2}, and
    the upper thin line corresponds to ROSAT/PSPC (\gl\gl 6--100~\AA)
    observations of a sample of M stars \citep{mar00}. The middle,
    short, thin line correspond to ROSAT/PSPC observations of AD~Leo
    \citep{mar97}.}
  \label{variab}
\end{figure*}

The long duration of the observations of AD~Leo with EUVE, spanning
$\sim$45~d,  allows us
to make some statistical studies on the distribution of variability of
this dM star. All the observations unaffected by the dead spot were used for
such statistics, although the results are essentially the same
when the rest of the data are included in the analysis. Different time
binning of the 
light curves (300, 600, 1200, 3000, 6000~s) was tried with the aim 
to explore different time scales.
In order to charachterize the variability of the star we plot a
distribution of the number of bins corresponding to 
the observed count rate (Fig.~\ref{variab}). The quiescent count rate
was found to be L$_{\rm qsc}$=0.20 cts~s$^{-1}$, with some points having
lower flux  
corresponding to the expected statistical deviations predicted by
fluctuations (a gaussian distribution with a width of
$\sigma$=$\sqrt{bin\times max}$~$bin^{-1}$, where {\it bin} is the binning
time and {\it max} is the maximum value).  The quiescent count rate was
then used to evaluate the 
cumulative variability amplitude (number of points with count rate above
a given value, defined as W=L/L$_{\rm qsc}$). An increase in the bin
size used in the light curves sampling (from 300 to 6000~s) results in
a slightly faster 
decrease of the distribution for larger amplitudes. This is the
consequence of smoothing the flaring behaviour in a larger bin size. 
The distribution of the amplitude variability using a 600~s
binning, shows a $\sim$W$^{-2.391\pm 0.006}$ distribution.
A similar distribution has been obtained for a 600~s binning of 
the light curves of YY~Gem (dM1/dM1) and FK~Aqr (dM2/dM3), as reported in
\citet{paper2}, spanning a total of $\sim$10~d and 19~d respectively,
as shown in Fig.~\ref{variab}. The time spanned for these two
objects is smaller than for AD~Leo, and this could hide effects
occuring at longer time scales, apart from being less representative of the
average behaviour of the mentioned stars.  Despite this caveat,
FK~Aqr shows a behaviour very similar to that of AD~Leo (with a dependency of
$\sim$W$^{-2.50\pm 0.01}$). 
On the other hand, YY~Gem had a lower number of flares,
resulting in a faster decrease for W$\la$3 ($\sim$W$^{-3.57\pm
  0.05}$), but for large flares it shows the same 
level of amplitude as AD~Leo.
For comparison we have reported also the distribution of
variability of a sample of dM~stars observed with ROSAT/PSPC
(6--100~\AA) by \citet{mar00}, and of the 15~ks ROSAT/PSPC observations of
AD~Leo \citep{mar97}. The flux detected at shorter 
wavelengths is expected to reflect the behaviour of material emitting
at higher temperature, that is more subject to flaring activity, producing
a the lower slope ($\sim$W$^{-1.69\pm
  0.02}$) in the distribution of the X-rays
observations of M stars. Although the 15~ks of ROSAT observations 
available (against 1.1~Ms of EUVE) are a short interval of time, 
the distribution of amplitude of AD~Leo in the 
ROSAT band (partially overlaping EUVE/DS) is remarkably similar to
that of EUVE.

\subsection{Spectra}

The spectrum of AD~Leo (Fig.~\ref{adleospec}) is dominated by lines
formed at high temperatures, like \ion{Fe}{xv}--{\sc xxiv}.
The good statistics achieved also permitted observing lines from
lower ionization states of Fe (\ion{Fe}{ix}--{\sc xiv}) and even three
\ion{Ne}{viii} lines, as listed in Table~\ref{euveflux}.
Finally, the frequent flaring behaviour of AD~Leo has 
allowed a good quality separated analysis of flaring and
quiescent spectra to be performed. Such analysis has shown 
an enhancement of most of the spectral
lines (see Fig.~\ref{adleospec}), and in particular the
hottest lines observable with EUVE, like Fe~{\sc xxiii}/{\sc xx}
$\lambda$133 and \ion{Fe}{xxiv} $\lambda$192, with a factor of $\sim$4 
and $\sim$15 (estimated from the blend with nearby lines, see
Table~\ref{euveflux}) increase between quiescent and flaring stages.
These enhancements affecting the hot lines
are larger than those found in the large flares observed by
\citet{paper1} in RS~CVn stars (binary systems usually including at least an
evolved late type star, and with high activity levels due to their
high rotation) like UX~Ari  or
$\sigma$~Gem, probably  due to the more impulsive 
behaviour of the AD~Leo flares. Note that the whole flaring process is 
averaged in these spectra, and hence a longer decay, like those
typically observed in the mentioned active binaries,  will decrease the
intensity of the spectral lines with respect to the peak of the flare.
An alternative explanation for these enhancements could be the
  hotter temperature of the flares 
in AD~Leo as compared to those in the RS~CVn stars.
However, the analysis of the temperature structure by means of the EMD
(see below) does not show any substantial difference during AD~Leo flares
in comparison with the flaring EMDs
reported in \citet{paper1}, and hence we can reject this possibility.

\begin{figure}
  \resizebox{\hsize}{!}{\includegraphics{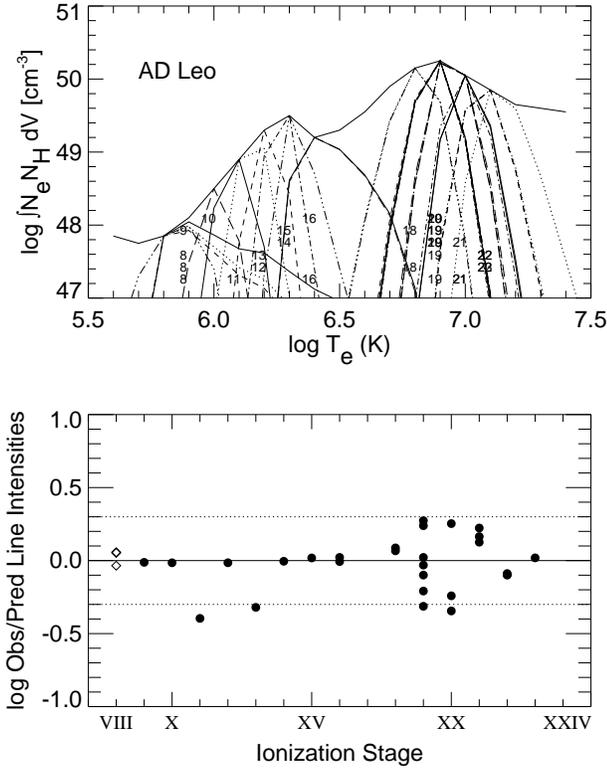}}
  \caption{{\it Upper}: EMD for the summed EUVE 
spectrum for AD Leo. Thin lines 
represent the relative contribution function for each ion (the 
emissivity function multiplied by the EMD at each point).
{\it Lower}: Observed to predicted line ratios for the ion 
stages in top figure. The dotted lines
denote a factor of 2. Filled circles are used for Fe 
ions, and diamonds for the three Ne lines.  
}
\label{adleoemd}
\end{figure}

The observed ratio of \ion{Fe}{xvi} $\lambda$$\lambda$ 335.41/360.80
from the summed spectra has been used to calculate the interstellar
hydrogen absorption. Its comparison with the theoretical branch ratio
yields a value $N_H$=3.4$^{+1.2}_{-1.5}\times$10$^{18}$~cm$^{-2}$, 
used in combination with a ratio 
\ion{He}{i}/\ion{H}{i}=0.09 \citep{kimb93} to correct the observed
fluxes for interstellar hydrogen and helium continuum absorption.
This value is consistent with the amount 
$N_H$=3$\times$10$^{18}$~cm$^{-2}$ reported by \citet{cul97} from the 
EUVE 1993 observations. Procyon, a nearby star to AD~Leo,
provides a value that can be checked for consistency.
\citet{lin95} obtained a close result 
($N_H$=1.2$\times$10$^{18}$~cm$^{-2}$) in Procyon by fitting the
Lyman~$\alpha$ profile.  

\begin{figure}
  \resizebox{\hsize}{!}{\includegraphics{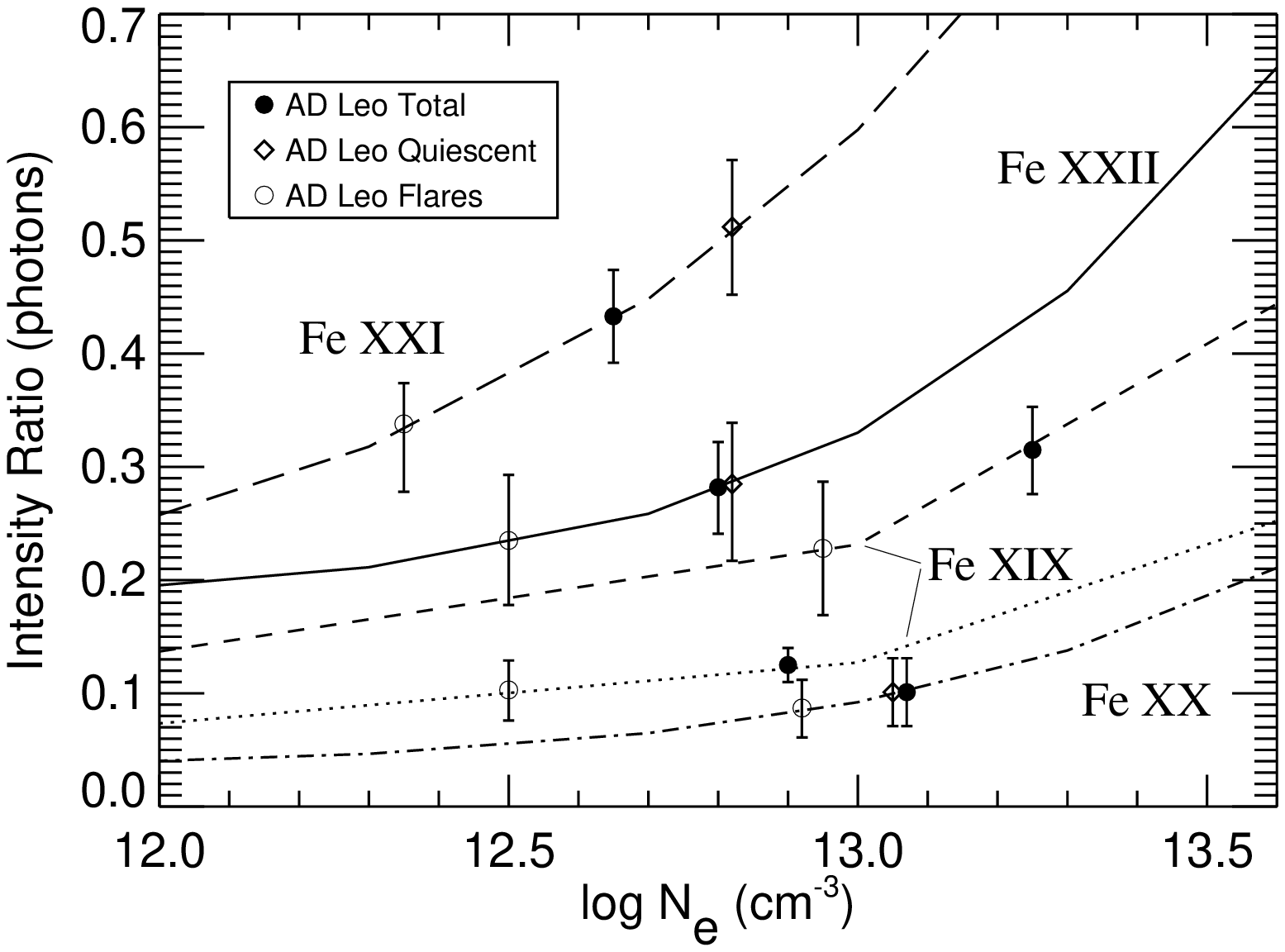}}
  \caption{Line ratios (photons) as a function of electron density
  ($N_e$) from the summed spectra.
  Ratios shown (from top to bottom) are:
  \ion{Fe}{xxi}~(\gl142.16+\gl 142.27)/\gl102.22,
  \ion{Fe}{xxi}~\gl102.22/\gl128.73,
  \ion{Fe}{xxii}~\gl114.41/\gl117.17,
  \ion{Fe}{xix}~\gl91.02/(\gl101.55+\gl109.97+\gl111.70),
  \ion{Fe}{xix}~\gl91.02/ (\gl108.37+\gl120.00), and
  \ion{Fe}{xx} \gl110.63/(\gl118.66+\gl121.83). The observed
  line ratios are plotted on the theoretical curves with 1$\sigma$
  error bars, following \citet{bri98}.  The flux for
  \ion{Fe}{xix} \gl91.02 is obtained from 
the blend with \ion{Fe}{xxi} using the \ion{Fe}{xxi} branching
ratio for \gl91.28 and \gl102.22.}
  \label{elecdens}
\end{figure}

Some electron density diagnostics are available from the observed
EUVE spectra. Line ratios from \ion{Fe}{xix}--{\sc xxii} have been
employed to calculate the electron density of the plasma emitting at
log~T(K)$\sim$7.0, resulting in an average value of log~$N_e$(cm$^{-3}$)
$\sim$12.9, 12.9 and 12.7 for
the summed, quiescent and flaring spectra respectively (see
Fig.~\ref{elecdens}). Although the 
data show some dispersion in the values, and in most cases the error
bars make the values of quiescent and flaring stages compatible. There
seems to be present a
tendency towards a decrease in the electron density during the flares. 
This behavior contrasts with the increase observed in some RS~CVn
stars analyzed by \citet{sanz01,paper1}. 
Further comments on this issue will follow in Sect.~\ref{sec:discussion}.

\begin{table}
\caption{\scshape Emission Measure Values for the Summed (S), Quiescent
  (Q) and Flaring (F) spectra}\label{tableemd}
\begin{center}
\begin{tabular}{lccc}
\hline \hline
{log T} & \multicolumn{3}{c}{log $\int N_e N_H\ dV$ (cm$^{-3}$)$^a$} \\
\cline{2-4}
{($K$)} & {S} & {Q} & {F} \\
\hline
 5.5 & 47.90: & 47.90: & 47.90: \\
 5.6 & 47.85: & 47.85: & 47.85: \\
 5.7 & 47.75: & 47.75: & 47.75: \\
 5.8 & 47.85 & 47.85 & 47.85 \\
 5.9 & 48.10 & 48.15 & 48.10 \\
 6.0 & 48.50 & 48.45 & 48.50 \\
 6.1 & 48.90 & 48.80 & 48.90 \\
 6.2 & 49.30 & 49.10 & 49.35 \\
 6.3 & 49.50 & 49.45 & 49.80 \\
 6.4 & 49.20 & 49.15 & 49.20 \\
 6.5 & 49.30 & 49.25 & 49.30 \\
 6.6 & 49.55 & 49.55 & 49.55 \\
 6.7 & 49.90 & 49.90 & 49.90 \\
 6.8 & 50.15 & 50.10 & 50.20 \\
 6.9 & 50.25 & 50.20 & 50.60 \\
 7.0 & 50.05 & 49.95 & 50.10 \\
 7.1 & 49.85 & 49.50 & 50.20 \\
 7.2 & 49.65 & 49.30 & 50.25 \\
 7.3 & 49.60 & 49.25 & 50.50 \\
 7.4 & 49.55 & 49.20 & 50.65 \\
 7.5 & 49.50: & 49.20: & 50.65: \\
 7.6 & 49.50: & 49.20: & 50.65: \\
 7.7 & 49.50: & 49.20: & 50.65: \\
 7.8 & 49.50: & 49.20: & 50.65: \\
\hline
\end{tabular}
\end{center}

$^a$Emission Measure, where $N_e$ 
and $N_H$ are electron and hydrogen densities, in $cm^{-3}$.
A colon indicates that the EMD value is uncertain because few lines 
occur in the temperature region.
\end{table}

\subsection{Emission Measure Distribution}

\begin{figure}
  \resizebox{\hsize}{!}{\includegraphics{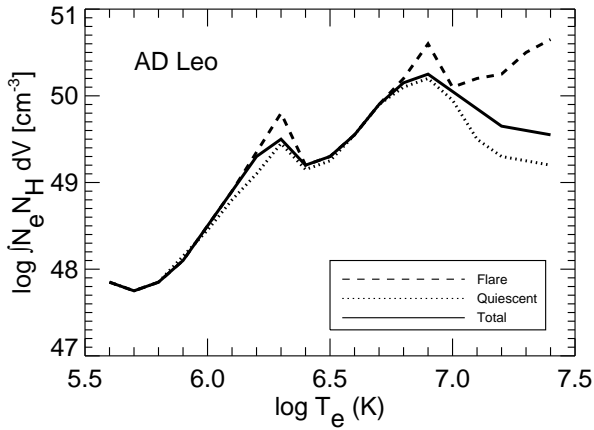}}
  \caption{Emission measure [$log \int N_eN_H\ dV$ (cm$^{-3}$)]
as a function of $log\ T_e$(K) for  AD~Leo during 
summed (solid line), flaring (dashed line) and quiescent (dotted 
line) stages.  The EMD above log T(K) = 7.4 is uncertain (see
text). }
  \label{stages}
\end{figure}

A line-based analysis of the emission spectra is employed in 
order to calculate the EMD ($\int N_e N_H dV$ cm$^{-3}$, where $N_e$
and $N_H$ are electron  and hydrogen densities, in cm$^{-3}$) of the
observed fluxes (Fig.~\ref{adleoemd}). The EUVE spectral coverage
permits using lines of 
all stages of iron ionization from \ion{Fe}{ix} through \ion{Fe}{xxiv}
except for \ion{Fe}{xvii}, which has no transitions in the EUV spectral
range. The analysis uses the line emissivities calculated by 
\citet{bri95} for the EUVE iron lines, based on a solar iron
abundance\footnote{The solar iron abundance is defined as (12. $+$
log~$Fe \over H$), where $Fe \over H$ represents the ratio of iron to 
hydrogen by number.} of 7.6 \citep{allen}, and corrected to match an
iron abundance of 7.67 \citep{anders}. Theoretical fluxes were
calculated using assumed EMDs \citep[see][and
references therein]{dup93,bri98,sanzcs12} which were then compared with the 
observed fluxes, in order to obtain the EMD that best fits the
fluxes within a factor of two.
Non iron lines, including some involved in complex blends,  have been
evaluated by using the Astrophysical Plasma  
Emission Code (APEC) v1.10 \citep{smith01}. 
The quiescent and flaring spectra have been used to perform a
separated analysis of their EMD (see Table~\ref{tableemd}). 
The flaring fluxes resulted in an 
increase of the EMD, progressively higher for higher temperatures, 
as displayed in Fig.~\ref{stages}. 
A similar analysis was made by \citet{paper1} for a sample of active
binary systems, resulting in 
a relatively uniform increase of the EMD in the range
log~T(K)$\sim$5.8--7.3. This contrasts with the behavior observed in
AD~Leo, where the EMD is more enhanced at hot temperatures (by a
factor of $\sim$9 at log~T(K)=7.2).

\begin{figure}
  \resizebox{\hsize}{!}{\includegraphics{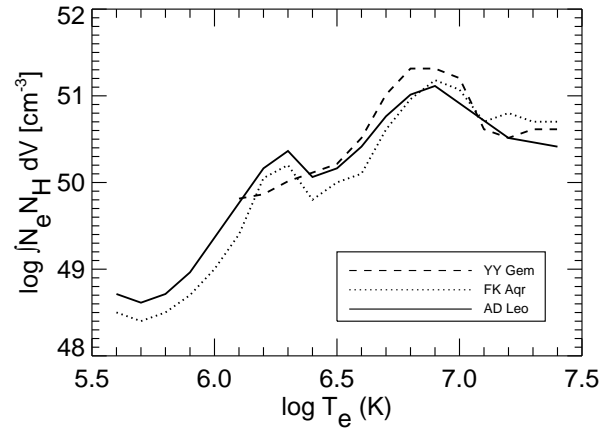}}
  \caption{Emission measure [$log \int N_eN_H\ dV$ (cm$^{-3}$)] as a
  function of $log\ T_e$(K) for  AD~Leo (dM3e), FK~Aqr (dM2e/dM3e) and 
  YY~Gem (dM1e/dM1e), weighted by their stellar radii in solar units. EMD of
  FK~Aqr and YY~Gem was taken from \citet{paper1}.}
  \label{m-stars}
\end{figure}

The EMD of AD~Leo has been compared with the EMD calculated in the
same manner by \citet{paper2} for the
active binary systems YY~Gem (dM1e/dM1e) and FK~Aqr (dM2e/dM3e)
weighted by the stellar radii as shown in Fig.~\ref{m-stars}.
The fast rotator YY~Gem (P$_{\rm phot}$=0.8~d) seems to have a slightly
larger EMD  
than AD~Leo (P$_{\rm phot}$=2.7~d) and FK~Aqr (P$_{\rm phot}$=4.4~d), although
an accurate knowledge of the elements abundances (specially Fe) would
be required to know the general level of each star. However,  the
shape remains very similar in the temperature range where the EMD is
well defined for the 3 stars, and it is almost coincident for FK~Aqr
and AD~Leo, for which there are data in the whole temperature range. 

\begin{figure}
  \resizebox{\hsize}{!}{\includegraphics{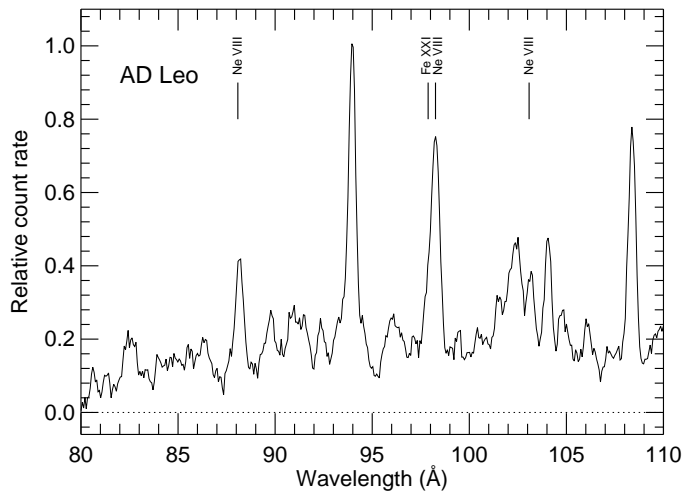}}
  \caption{Close-up of the EUVE SW summed spectrum with the spectral
    region containing the three 
    \ion{Ne}{viii} lines involved in the calculation of Ne abundance.}
  \label{nelines}
\end{figure}

Once the EMD was calculated using only iron lines, three \ion{Ne}{viii}
lines (see Table~\ref{euveflux} and Fig.~\ref{nelines}) 
were added to the analysis in order to estimate the abundance of
Ne based on \citet{anders} solar photospheric abundances. 
The lines of \ion{Fe}{ix} and \ion{Fe}{x} observed in the EUVE
spectra permit us to have a reference of the EMD at log~T(K)$\sim$5.9,
the same temperature range at which the \ion{Ne}{viii} lines are formed (see
Fig.~\ref{adleoemd}), with error bars that can be estimated directly
from the observational uncertainties in the 3 Ne and 2 Fe lines
involved in the analysis.
Once this abundance is set to [Ne/Fe]=1.05$\pm$0.08,
the predicted lines have shown a fit within 0.1~dex with the EMD
predicted from EUVE. 
The abundance found in the quiescent and flaring phases was
[Ne/Fe]=1.00$\pm$0.09 and [Ne/Fe]=1.15$\pm$0.18 respectively. Although
there could be some increment in the value of the abundance of Ne
during flares with respect to quiescence, the
uncertainties in the calculation prevent any conclusion. The value of
abundance of Ne found is 
inconsistent with the value of 0.4 calculated by 
\citet{mag01} from Chandra/LETG data. 
Some
factors could be responsible for the discrepancy: although based on a
smaller number of lines, the EUVE spectra allow us to cover
the whole temperature range log~T(K)$\sim$5.8--7.3 using only Fe
lines; as a consequence, the two lines of \ion{Fe}{ix} and
\ion{Fe}{x} can be used in combination with the
\ion{Ne}{viii} lines formed at the same temperature. Chandra spectra
instead, permit using more lines, though combined from different
elements in the whole   
temperature range; the uncertainties in the knowledge of abundances of the
elements may influence the result, as well as a worse
temperature coverage for Fe lines alone, used as reference for the
rest of the analysis in \citet{mag01}. The exact shape of the EMD may
strongly influence the analysis of the abundance found from
Chandra spectra, where some lines of \ion{Ne}{x} showed some
discrepancy with the EMD and the abundances proposed by
\citet{mag01}, while the presence of Fe lines in the same
temperature range makes the EUVE analysis independent of the shape
of the EMD. 

The large value of the Ne abundance found with EUVE data in
AD~Leo is remarkably similar to the value of [Ne/Fe]=0.99$\pm$0.11, found in
the active binary system 
V711~Tau from {\em Chandra}  High-Energy Transmission Grating data by 
\citet{dra01}. The EMD of this system is also 
similar to that of AD~Leo, up to log~T(K)$\sim$7.0, and this Ne
enhancements could be somehow related to the structures formed in the
range log~T(K)$\sim$6.4--6.9, not commonly observed in the Sun.

\section{Discussion}\label{sec:discussion}

The light curves and EMD of the flaring stages of AD~Leo show a
different behavior with respect to that observed in the RS~CVn stars
analyzed in \citet{paper1}:
(i) flares in AD~Leo and other M dwarfs (may be even G or K dwarfs)
tend to be short and intense, in contrast with the long-lasting flares
observed in giant and subgiant G or K stars; 
(ii) the EMD is largely enhanced at the hottest
temperatures during flares, differently from the uniform increase of
the EMD observed in the active binaries with an evolved component; 
(iii) finally, the electron density at log~T(K)$\sim$6.9 on the
evolved stars 
increases during flares, while in AD~Leo the electron
density, at most, remains constant.

Previous work by \citet{cul97} calculated the EMD of AD~Leo for
different activity levels of the 1993 observations. Although poorly
constrained due to low statistics, the authors found
a quiescent EMD dominated by material at log~T(K)$\sim$6.8--7.2 that
they interpreted as a distribution of coronal loops with peak
temperatures in that range. The application of loop models
developed by \citet{haw95} led \citet{cul97} to estimate long loop
lengths and densities in the range log~$N_e$(cm$^{-3}$)$\sim$9--11
at flare peak.

The EMD calculated in the present paper could be, in principle,
interpreted as the combination 
of different families of loops peaking at temperatures of
log~T(K)$\sim$6.3, log~T(K)$\sim$6.9 and  log~T(K)$\ga$7.3, as
proposed in \citet{paper1}. 
According to that explanation, the material emitting at
log~T(K)$\sim$6.9 would actually correspond to the loops of the second
and third families of loops. Hence, the electron density measured at
that temperature is weighting the number of loops of both families,
families that could actually be characterized by different
densities. Some of these loops could actually be related with
unresolved solar-like flares, as proposed by
\citet{rea01}, and this 
would explain the high electron densities measured in these stars even
in the absence of large flares.
The observed tendency towards a decrease in electron
density in AD~Leo, rather than being real might be related to the
mentioned balance of loops. 
When the observed flares occurred, the number of the third
group of loops would increase, while the number of loops of the
first and second groups would not grow so much, or they could
be transforming into hotter loops. If the third group was made of
flaring loops with lower density, the increase in the
number of these loops would result in a decrease in the measured
density at log~T(K)$\sim$6.9, unless there was sufficient increment in
density in the second group of loops. 
In the case of the active binary systems this
effect is not present because the number of loops of the third group
is already high even in the ``quiescent'' stage, and the
addition of such loops would be unable to compensate the possible
increment of density produced in the second family of loops.  
It must be noticed that 
the observed quiescent flux level
could actually be reflecting some flaring activity not
resolved by the time scale of the EUVE light curves. 

It is possible to have a first approximation, for the case of AD~Leo,
on the scale size of the third family of loops as
compared to those of the second group, 
by applying the scaling laws proposed by \citet{rtv78}:
$T^2=c n L$, where $T$ is the maximum
temperature of the loop, $c$ is a constant independent of the
loop characteristics, $n$ is the density in the loop, and $L$ is the
loop length 
scale. Hence, if we assume to have the same density in the
second and third group of loops (it could even be lower in the third
group), an increment in temperature would 
automatically imply a larger loop length, and hence we can
expect that the loops peaking at log~T(K)$\ga$7.1 will have a larger
scale size than those peaking at log~T(K)$\sim$6.9.

The differences observed between the flares in AD~Leo (and other dM
stars, may be even G and K dwarfs) and those in some RS~CVn
systems, seem to point to a different flaring mechanism in M dwarf
stars with respect to that of evolved
earlier type stars. While the latter can
sustain a long-lasting flare, with duration of several days in the EUV
and X-rays bands, the dM stars seem to have short duration flares,
with faster decays after the peak has been reached, and affecting
especially the region at log~T(K)$\ga$7.1. A continued heating or a
different loop length could be directly responsible for these
differences. The different stellar structure of dwarf and giant (and
subgiant) stars may be the physical reason behind these different
flaring behaviors.

\section{Conclusions}\label{sec:conclusions}
We have analyzed 1.1~Ms of EUVE observations of AD~Leo. The star shows
a relevant flaring activity allowing us to study the variability
frequency in the EUV band and the spectral properties during both the
quiescent and the flaring states. Results imply the next conclusions:

\begin{itemize}

\item Statistics carried out on flaring frequency show similar
  behavior to other active dM stars like FK~Aqr, with a well defined
  quiescent stage and a distribution of
  amplitude variability following a power law with an index of $\sim -2.4$.

\item The Emission Measure Distribution (EMD) of AD~Leo has been calculated
  from EUVE spectra. The EMD is dominated by a
  well defined enhancement peaking around
  log~T(K)$\sim$6.9, and shows the presence of a second peak at
  log~T(K)$\sim$6.3. The enhancements remained constant in temperature
  regardless of flaring activity.

\item A large increase in the amount of material at log~T(K)$\ga$7.1 has been
  found during flaring stages, with an EMD increment higher than at
  lower temperatures. The electron density at log~T(K)$\sim$7
  [log~$N_e$(cm$^{-3}$)$\sim$12.9] did not show any enhancement during flares,
  contrary to the tendency observed in some RS~CVn stars.

\item The combination of three families of loops, with maximum
  temperatures at log~T(K)$\sim$6.3, log~T(K)$\sim$6.9 and somewhere
  at log~T(K)$\ga$7.1 respectively, has been proposed to explain the observed
  EMD. Electron density in the loops peaking at log~T(K)$\ga$7.1
  must be lower than the density of loops at log~T(K)$\sim$6.9. 
  Differences in the stellar structure of dwarf and evolved stars
  are the most probable cause for the differences found in their flaring
  behavior. 

\item A large value of the abundance of Ne
  ([Ne/Fe]$\sim$1.05$\pm$0.08) at log~T(K)$\sim$5.9 was found with
  respect to solar 
  photospheric value. No significative increment in the abundance was
  detected during flares. 

\end{itemize}

\begin{acknowledgements}

We acknowledge support by the Marie Curie Fellowships Contract No.
HPMD-CT-2000-00013. 
We have made use of data obtained through 
the High Energy Astrophysics Science Archive Research Center Online
Service, provided by the NASA/Goddard Space Flight Center.  This
research has also made use of NASA's Astrophysics Data System Abstract
Service.
\end{acknowledgements}


\begin{thebibliography}{28}
\expandafter\ifx\csname natexlab\endcsname\relax\def\natexlab#1{#1}\fi

\bibitem[{{Allen}(1973)}]{allen}
{Allen}, C.~W. 1973, Astrophysical Quantities (London: University of London,
  Athlone Press, |c1973, 3rd ed.)

\bibitem[{{Anders} \& {Grevesse}(1989)}]{anders}
{Anders}, E. \& {Grevesse}, N. 1989, \gca, 53, 197

\bibitem[{{Brickhouse} \& {Dupree}(1998)}]{bri98}
{Brickhouse}, N.~S. \& {Dupree}, A.~K. 1998, \apj, 502, 918

\bibitem[{{Brickhouse} {et~al.}(1995){Brickhouse}, {Raymond}, \&
  {Smith}}]{bri95}
{Brickhouse}, N.~S., {Raymond}, J.~C., \& {Smith}, B.~W. 1995, \apjs, 97, 551

\bibitem[{{Covino} {et~al.}(2001){Covino}, {Panzera}, {Tagliaferri}, \&
  {Pallavicini}}]{cov01}
{Covino}, S., {Panzera}, M.~R., {Tagliaferri}, G., \& {Pallavicini}, R. 2001,
  \aap, 371, 973

\bibitem[{{Cully} {et~al.}(1997){Cully}, {Fisher}, {Hawley}, \&
  {Simon}}]{cul97}
{Cully}, S.~L., {Fisher}, G.~H., {Hawley}, S.~L., \& {Simon}, T. 1997, \apj,
  491, 910

\bibitem[{{Drake} {et~al.}(2001){Drake}, {Brickhouse}, {Kashyap}, {Laming},
  {Huenemoerder}, {Smith}, \& {Wargelin}}]{dra01}
{Drake}, J.~J., {Brickhouse}, N.~S., {Kashyap}, V., {et~al.} 2001, \apjl, 548,
  L81

\bibitem[{{Dupree} {et~al.}(1993){Dupree}, {Brickhouse}, {Doschek}, {Green}, \&
  {Raymond}}]{dup93}
{Dupree}, A.~K., {Brickhouse}, N.~S., {Doschek}, G.~A., {Green}, J.~C., \&
  {Raymond}, J.~C. 1993, \apjl, 418, L41

\bibitem[{{Favata} {et~al.}(2000){Favata}, {Micela}, \& {Reale}}]{fav00}
{Favata}, F., {Micela}, G., \& {Reale}, F. 2000, \aap, 354, 1021

\bibitem[{{G{\" u}del} {et~al.}(2001){G{\" u}del}, {Audard}, {Guinan}, {Mewe},
  {Drake}, \& {Alekseev}}]{gud01}
{G{\" u}del}, M., {Audard}, M., {Guinan}, E.~F., {et~al.} 2001, in ASP
  Conf.~Ser. 223, The Eleventh Cambridge Workshop on Cool Stars, Stellar
  Systems and the Sun, eds. R.\ J. Garc\'{\i}a L\'opez, R. Rebolo, \& M.\ R.
  Zapatero Osorio (San Francisco: ASP), 1085

\bibitem[{{Hawley}(2001)}]{haw01}
{Hawley}, S.~L. 2001, in ASP Conf.~Ser., The Twelfth Cambridge Workshop on
  Cool Stars, Stellar Systems and the Sun, eds. T. Ayres \& A. Brown (Boulder:
  ASP)

\bibitem[{{Hawley} {et~al.}(1995){Hawley}, {Fisher}, {Simon}, {Cully},
  {Deustua}, {Jablonski}, {Johns-Krull}, {Pettersen}, {Smith}, {Spiesman}, \&
  {Valenti}}]{haw95}
{Hawley}, S.~L., {Fisher}, G.~H., {Simon}, T., {et~al.} 1995, \apj, 453, 464

\bibitem[{{Henry} {et~al.}(1994){Henry}, {Kirkpatrick}, \& {Simons}}]{hen94}
{Henry}, T.~J., {Kirkpatrick}, J.~D., \& {Simons}, D.~A. 1994, \aj, 108, 1437

\bibitem[{{Kimble} {et~al.}(1993){Kimble}, {Davidsen}, {Long}, \&
  {Feldman}}]{kimb93}
{Kimble}, R.~A., {Davidsen}, A.~F., {Long}, K.~S., \& {Feldman}, P.~D. 1993,
  \apjl, 408, L41

\bibitem[{{Linsky} {et~al.}(1995){Linsky}, {Diplas}, {Wood}, {Brown}, {Ayres},
  \& {Savage}}]{lin95}
{Linsky}, J.~L., {Diplas}, A., {Wood}, B.~E., {et~al.} 1995, \apj, 451, 335

\bibitem[{{Maggio} {et~al.}(2001){Maggio}, {Drake}, {Kashyap}, {Micela},
  {Sciortino}, {Peres}, {Harnden}, \& {Murray}}]{mag01}
{Maggio}, A., {Drake}, J.~J., {Kashyap}, V., {et~al.} 2001, in Stellar Coronae
  in the Chandra and XMM-Newton era, ed. by J. Drake, \& F. Favata (Noordwijk:
  ASP), in press

\bibitem[{{Marino}(1997)}]{mar97}
{Marino}, A. 1997, Tesi di Laurea, University of Palermo

\bibitem[{{Marino} {et~al.}(2000){Marino}, {Micela}, \& {Peres}}]{mar00}
{Marino}, A., {Micela}, G., \& {Peres}, G. 2000, \aap, 353, 177

\bibitem[{{Miller-Bagwell} \& {Abbott}(1995)}]{mill95}
{Miller-Bagwell}, A. \& {Abbott}, M. 1995, EUVE Guest Observer Data Products
  Guide

\bibitem[{{Osten} {et~al.}(2000){Osten}, {Brown}, {Ayres}, {Linsky}, {Drake},
  {Gagn{\' e}}, \& {Stern}}]{ost00}
{Osten}, R.~A., {Brown}, A., {Ayres}, T.~R., {et~al.} 2000, \apj, 544, 953

\bibitem[{{Reale} {et~al.}(2001){Reale}, {Peres}, \& S.}]{rea01}
{Reale}, F., {Peres}, G., \& S., O. 2001, \apj, in press

\bibitem[{{Rosner} {et~al.}(1978){Rosner}, {Tucker}, \& {Vaiana}}]{rtv78}
{Rosner}, R., {Tucker}, W.~H., \& {Vaiana}, G.~S. 1978, \apj, 220, 643

\bibitem[{{Sanz-Forcada}(2001)}]{sanzcs12}
{Sanz-Forcada}, J. 2001, in ASP Conf.~Ser., The Twelfth Cambridge Workshop
  on Cool Stars, Stellar Systems and the Sun, eds. T. Ayres \& A. Brown
  (Boulder: ASP), in press

\bibitem[{{Sanz-Forcada} {et~al.}(2001){Sanz-Forcada}, {Brickhouse}, \&
  {Dupree}}]{sanz01}
{Sanz-Forcada}, J., {Brickhouse}, N.~S., \& {Dupree}, A.~K. 2001, \apj, 554,
  1079

\bibitem[{{Sanz-Forcada} {et~al.}(2002{\natexlab{a}}){Sanz-Forcada},
  {Brickhouse}, \& {Dupree}}]{paper1}
---. 2002{\natexlab{a}}, \apj, 570, 799

\bibitem[{{Sanz-Forcada} {et~al.}(2002{\natexlab{b}}){Sanz-Forcada},
  {Brickhouse}, \& {Dupree}}]{paper2}
---. 2002{\natexlab{b}}, \apj, submitted

\bibitem[{{Smith} {et~al.}(2001){Smith}, {Brickhouse}, {Liedahl}, \&
  {Raymond}}]{smith01}
{Smith}, R.~K., {Brickhouse}, N.~S., {Liedahl}, D.~A., \& {Raymond}, J.~C.
  2001, \apjl, 556, L91

\bibitem[{{Spiesman} \& {Hawley}(1986)}]{spi86}
{Spiesman}, W.~J. \& {Hawley}, S.~L. 1986, \aj, 92, 664

\end{thebibliography}
\end{document}